\documentclass[10pt,conference]{IEEEtran}

\usepackage{physics}
\usepackage{graphicx}% Include figure files
\usepackage{dcolumn}% Align table columns on decimal point
\usepackage{booktabs}
\usepackage{bm}% bold math
\usepackage{makecell}
\usepackage{tabularx}
\usepackage{xcolor}
\usepackage{algorithm}
\usepackage{algpseudocode}
\usepackage{hyperref}
\usepackage{cite}
\usepackage[utf8]{inputenc}
\hypersetup{
   breaklinks=true,   % splits links across lines
   colorlinks=true,   % displays links as colored text instead of blocks
   citecolor=magenta,
   linkcolor=magenta,
   urlcolor=magenta
}

\newcommand{\param}{\vec \alpha}

\DeclareUnicodeCharacter{2009}{ }

\begin{document}

\title{Efficient control pulses for continuous quantum gate families through coordinated re-optimization
\thanks{Identify applicable funding agency here. If none, delete this.}
}

\author{\IEEEauthorblockN{Jason D. Chadwick}
\IEEEauthorblockA{\textit{Department of Computer Science} \\
\textit{University of Chicago}\\
Chicago, IL, USA \\
jchadwick@uchicago.edu}
\and

\IEEEauthorblockN{Frederic T. Chong}
\IEEEauthorblockA{\textit{Department of Computer Science} \\
\textit{University of Chicago}\\
Chicago, IL, USA \\
chong@cs.uchicago.edu}}

\maketitle

\begin{abstract}
We present a general method to quickly generate high-fidelity control pulses for any continuously-parameterized set of quantum gates after calibrating a small number of reference pulses. We find that interpolating between optimized control pulses for different quantum operations does not immediately yield a high-fidelity intermediate operation. To solve this problem, we propose a method to optimize control pulses specifically to provide good interpolations. We pick several reference operations in the gate family of interest and optimize pulses that implement these operations, then iteratively re-optimize the pulses to guide their shapes to be similar for operations that are closely related. Once this set of reference pulses is calibrated, we can use a straightforward linear interpolation method to instantly obtain high-fidelity pulses for arbitrary gates in the continuous operation space. 

We demonstrate this procedure on the three-parameter Cartan decomposition of two-qubit gates to obtain control pulses for any arbitrary two-qubit gate (up to single-qubit operations) with consistently high fidelity. Compared to previous neural network approaches, the method is 7.7x more computationally efficient to calibrate the pulse space for the set of all single-qubit gates. Our technique generalizes to any number of gate parameters and could easily be used with advanced pulse optimization algorithms to allow for better translation from simulation to experiment.
\end{abstract}

\section{Introduction}\label{sec:intro}

Quantum circuits, consisting of logical operations on qubits, are typically decomposed into a set of elementary basis operations that are specific to a given hardware device. These basis operations can be individually calibrated through control pulse shaping to achieve high accuracy. However, the space of all quantum operations is much larger than just the hardware basis set, meaning that operations must typically be decomposed into a sequence of basis operations. The capability to perform any operation within a continuous gate set could significantly improve the capabilities of near-term quantum computers, avoiding the runtime and fidelity costs associated with decomposing into basis operations. 

Continuous gate sets have been shown to improve fidelity and reduce gate count of quantum circuits \cite{lacroix_improving_2020, gokhale_optimized_2020, earnest_pulse-efficient_2021}. The implementation of continuous gate sets has been studied on various hardware platforms such as trapped ions \cite{sorensen_quantum_1999, sorensen_entanglement_2000} and various superconducting platforms \cite{foxen_demonstrating_2020, collodo_implementation_2020, gokhale_optimized_2020, earnest_pulse-efficient_2021}. However, each of these studies focused on implementing a specific continuous gate. In this work, we provide a hardware-agnostic method to calibrate any arbitrary continuous gate set. Our approach uses a quantum optimal control solver as a subroutine, iteratively optimizing control pulses to implement a few specific operations from within the gate set, from which we obtain pulses for any other operation.

\begin{figure}[t]
    \centering
    \includegraphics[width=\linewidth]{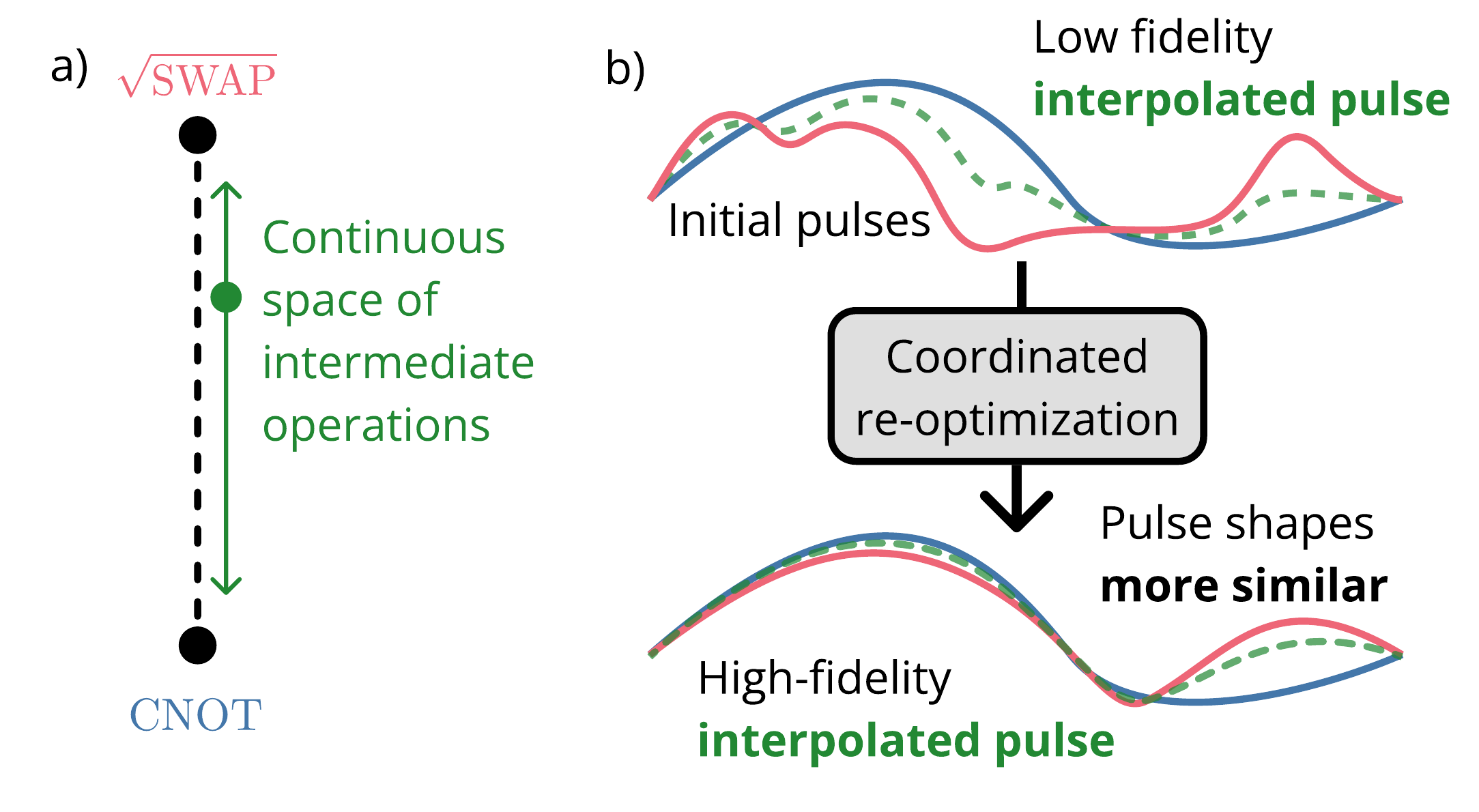}
    \caption{a) We seek to efficiently generate high-fidelity control pulses for continuous families of quantum gates. Here, we envision some continuous set of unitary operations between the $CNOT$ and $\sqrt{SWAP}$ operations. The goal is to efficiently obtain control pulses for any arbitrary point in this one-dimensional space of operations while only explicitly calibrating pulses at the two endpoints. b) \emph{Top:} Consider some high-fidelity control pulses that implement $CNOT$ and $\sqrt{SWAP}$ (blue and red). We attempt to obtain a pulse for an intermediate operation (green) through linear interpolation. We find that interpolation yields poor results when the fixed pulses have very different shapes. \emph{Bottom:} However, if we can re-optimize the pulses for $CNOT$ and $\sqrt{SWAP}$ to be more similar to each other (while still performing the correct operations), our simple linear interpolation method can obtain a high-fidelity pulse for the intermediate operation. Our methods generalize to higher-dimensional parameter spaces.}
    \label{fig:hero}
\end{figure}

Quantum control optimization involves shaping hardware control pulses to execute a target operation with high fidelity. Many software packages have been designed to solve quantum control problems for various systems and objectives such as \cite{khaneja_optimal_2005, doria_optimal_2011, caneva_chopped_2011, leung_speedup_2017, petersson_discrete_2020, gunther_quandary_2021, wu_data-driven_2018,  valahu_quantum_2022, goldschmidt_model_2022}. If the hardware device of interest is fully controllable, any unitary quantum operation can be implemented using an optimal control solver. Thus, a simple way to support arbitrary continuous gate sets through optimal control is to optimize pulses on an as-needed basis, depending on the operations encountered in the chosen circuit. However, pulse optimizations are computationally expensive, making it infeasible to do this when the input circuit is not necessarily known in advance and long delays in execution time are undesirable.

We address this problem by describing a procedure to pre-calibrate a continuous pulse landscape for a family of quantum operations, from which high-fidelity control pulses for arbitrary operations can be instantly retrieved. We create this landscape by picking a small number of specific operations to directly optimize pulses for, re-optimizing these pulses to be similar to one another, and then defining an interpolation function to retrieve new pulses for any operation in between.

\begin{figure*}[t]
    \centering
    \includegraphics[width=\linewidth]{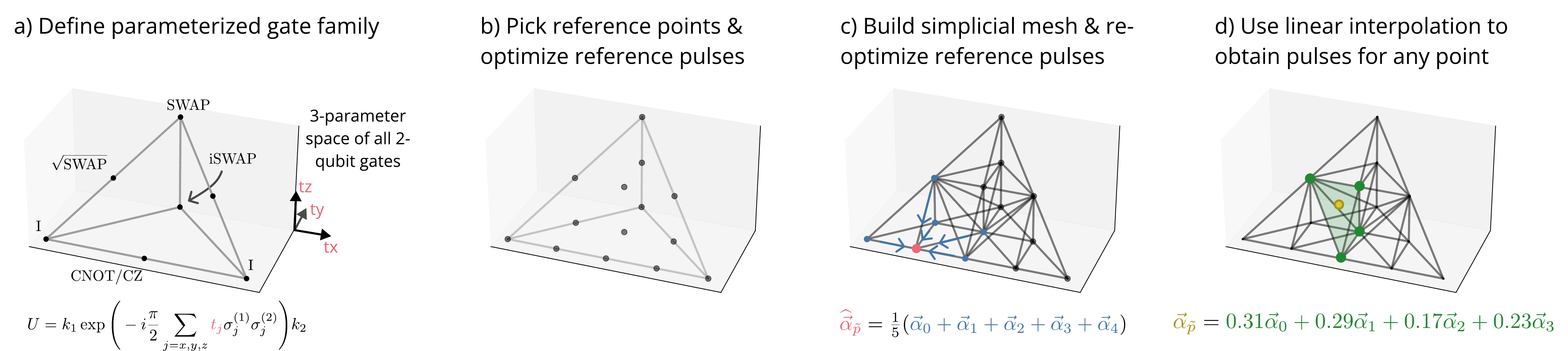}
    \caption{Overview of the example of our method provided in this work. More details on this specific example are provided in Section \ref{sec:evaluation}. a) We use the Cartan decomposition \eqref{eq:cartan} to define a three-dimensional space that contains all two-qubit operations. Our goal is to efficiently generate pulses for any arbitrary point within this continuous space. b) We choose several \emph{reference operations} within the space and generate control pulses that implement these operations via optimal control software. Interpolating between these reference pulses does not necessarily provide a high-quality pulse immediately. c) To solve this problem, we iteratively re-optimize each reference pulse to be more similar to the average of nearby reference pulses, a process we call \emph{neighbor-average re-optimization}. This process is described in Section \ref{sec:reference-pulses}. d) Once these reference pulses have been fully calibrated, we can instantly obtain control pulses for any operation in the continuous family by performing linear interpolation between the nearest reference pulses. This process is described in more detail in Section \ref{sec:interpolation}.}
    \label{fig:example}
\end{figure*}

The idea of applying optimal control to continuous spaces is not new. Reference \cite{luchi_control_2022} studied the problem of applying a specific operation on a parameterized Hamiltonian to create resilience to varying device errors. Reference \cite{shi_simulating_2021} examines the single-parameter case in the optimization of a cubic interaction unitary, performing interpolation between two fixed control pulses in a single-parameter space with consistently high fidelity. Reference \cite{li_optimal_2011} introduces methods to generate continuous sets of controls to create robustness to experimental deviations in Hamiltonians.

Reference \cite{sauvage_optimal_2022} (from which we borrow the term ``gate family'') demonstrates the effectiveness of a neural network for generating control pulses for families of parameterized gates. The network takes as inputs the parameters of a specific gate and the time value $t$, and it outputs the control value(s) $f(t)$. The network is trained over many iterations consisting of batches of randomly-sampled operations from the gate family. Reference \cite{preti_continuous_2022} uses a similar neural network method, but does not provide detailed data for larger parameter spaces with which to make an objective comparison, so we focus primarily on comparisons with \cite{sauvage_optimal_2022} in this work.

% In the examples presented in the text, the network training cost is 51,200 total cost function evaluations. While this can be considered a reasonably small number when simulating pulses on computers, it can quickly become expensive for an implementation on physical devices, considering that the network would likely need to be recalibrated frequently to account for device variations over time \cite{murali_noise-adaptive_2019, tannu_not_2019} and that the network will not have access to a perfect device model, instead needing to rely on imperfect and incomplete state measurements.

Figure \ref{fig:hero}a shows an example of the problem we solve. We start with two \emph{reference} operations, $CNOT$ and $\sqrt{SWAP}$, with the goal of efficiently obtaining pulses for any intermediate through interpolation. Under the assumption that sufficiently small changes in control pulse values will lead to small changes in the realized quantum operation, linear interpolation between control pulses should yield reasonable results if the initial pulses are similar enough. However, in general, the same quantum operation can be achieved with many different physical pulse sequences; an optimal control solver usually starts from random initial guesses, and, unless heavily constrained, is not expected to reach the same final pulse shape for different initial guesses. Two independently-optimized pulse sequences for different quantum operations could therefore have extremely different pulse shapes (even if the operations themselves are quite similar), meaning that interpolation will generally not provide good results, as observed in the Supplemental Material of \cite{sauvage_optimal_2022}. 

Figure \ref{fig:hero}b shows the key insight of this work: while optimizing pulses for the two reference operations initially yields poor interpolation accuracy, re-optimizing these pulses to be more similar to each other constrains the pulse solution space and provides significant improvements in average interpolation fidelity. Once the reference pulses are optimized, interpolation can instantly retrieve the pulse for any operation in the parameter space.

In this work, we describe \emph{coordinated re-optimization}, a method to iteratively optimize reference pulses to be more similar, which enables the use of interpolation to efficiently generate control pulses for continuously-parameterized quantum operations. Our method can achieve equal or better average infidelities as the neural network methods of \cite{sauvage_optimal_2022} and \cite{preti_continuous_2022} using less computation. Additionally, any optimal control algorithm can be used in this framework, allowing for advanced techniques for robust or closed-loop pulse control such as \cite{leung_robust_2018, valahu_quantum_2022, wu_data-driven_2018, brown_arbitrarily_2004, goldschmidt_model_2022} to directly replace or augment the optimal control unit in our method; these pulse optimization algorithms could significantly improve the translation between simulation and experiment.

We specifically present applications of our method to the set of all single-qubit gates and to the set of two-qubit gate unitaries, which are parameterized by Weyl chamber coordinates, under the belief that these classes of gates are the most important for enabling near-term applications on quantum computers. However, our method can be generalized to any parameterized gate family.

\section{Optimizing reference pulses}\label{sec:reference-pulses}

We use the Boulder Opal optimization package \cite{ball_software_2021, q-ctrl_boulder_2022}. We describe a control pulse by its vector of optimizable variables $\vec \alpha$. We use the objective function 
% \begin{align} \label{eq:cost}
%     J = \underbrace{1 - \frac{1}{h^2} \abs{\text{Tr}\Big(U^\dagger_{\text{target}}U_T\Big)}^2}_{\text{gate infidelity}}
%     + \underbrace{\frac{\lambda}{n_f} \sum_k^{n_f}  \frac{\norm{\vec \alpha_k - \vec \alpha_{0,k}}_2^2}{n_p \cdot \alpha_{\text{max}}^2}.}_{\text{Tikhonov regularization}}
% \end{align}
\begin{align} \label{eq:cost}
    J = \underbrace{1 - \frac{1}{h^2} \abs{\text{Tr}\Big(U^\dagger_{\text{target}}U_T(\vec{\alpha})\Big)}^2}_{\text{gate infidelity}}
    + \underbrace{\widetilde\lambda \sum_k^{n_f}  \norm{\vec \alpha_k - \vec \alpha_{0,k}}_2^2,}_{\text{Tikhonov regularization}}
\end{align}
with Tikhonov weight
\begin{align}
    \widetilde \lambda = \frac{\lambda}{n_f \cdot n_p \cdot \alpha^2_{\text{max}}}.
\end{align}
Gate infidelity is calculated between the target unitary $U_{\text{target}}^\dagger$ and the unitary $U_T(\vec \alpha)$ implemented by the pulses. Two-qubit gates have Hilbert space dimension $h = 4$. The Tikhonov regularization term penalizes distance between $n_f$ optimizable pulses, each with $n_p$ optimization variables stored in the vector $\vec \alpha_k$, and fixed target vector $\vec \alpha_{0,k}$ with user-specified weight $\lambda$. It is common to include a similar regularization term in the cost function with $\param_0 = \vec 0$ to encourage low-amplitude control pulses, as in \cite{anders_petersson_optimal_2022}. In this work, we found $\lambda = 10^{-2}$ to yield satisfactory results, but this will likely require tuning for different devices and gate families. Choice of $\param_0$ is key to our method and is discussed in detail in Section \ref{sec:neighbor-avg}.

\subsection{Initial optimization}

The first step of our method is to parameterize the gate family of interest and define the volume in parameter space that contains all desired gates, as seen in Figure \ref{fig:example}a. To create the interpolation landscape, we first pick reference points on a rectangular grid in this parameter space, where the grid spacing is a tunable parameter that we refer to as the \emph{granularity}. For each point, we obtain the corresponding quantum operation and then use the pulse optimizer to find a valid control pulse sequence that implements this operation. For this initial round of optimization, we choose Tikhonov target $\param_0 = \vec 0$ in \eqref{eq:cost} to encourage low-amplitude pulses.

As can be seen in the examples in later sections, this initial pulse set does not always interpolate well in all parts of parameter space. There is no guarantee that control pulses for similar unitaries will themselves be similar, as observed in \cite{sauvage_optimal_2022}; there are often many distinct control sequences that can realize the same unitary operation, yielding near-useless interpolations in between different reference pulses. Tikhonov regularization with $\param_0 = \vec 0$ helps by restricting the space of allowed solutions, but does not fully resolve this issue. This motivates a more intelligent approach to finding reference pulses, which we call \emph{neighbor-average re-optimization}.

\subsection{Neighbor-average re-optimization} \label{sec:neighbor-avg}

We generate a simplicial mesh across our reference points using the \texttt{scipy.spatial.Delaunay} function \cite{virtanen_scipy_2020}. A $d$-simplex is the simplest $d$-dimensional polytope, formed by the convex hull of $d+1$ vertices. For example, a 2-simplex is a trangle and a 3-simplex is a tetrahedron. For a set of $k \geq d+1$ points in $d$ dimensions, a \emph{simplicial mesh} of contiguous $d$-simplices can be generated to cover the space between the points\footnote{Issues may arise in specific configurations of points. For example, if all points are coplanar in a 3D space, no nonzero-volume tetrahedra can be generated. However, in practice, we can reliably generate a simplicial mesh for any reasonable set of points.}. Each point is then the vertex of one or more simplices in the mesh. An example of a 3-dimensional simplicial (tetrahedral) mesh is shown in Figure \ref{fig:example}c. From this simplicial mesh, we can query the \emph{neighbors} of a given reference point, which are all other points that are connected to the point of interest by an edge.

Starting with the initial naive reference pulses from the initial optimization round, we generate a simplicial mesh over the reference points. For each reference point $p_i$, we find the set of neighboring vertices $\eta(p_i)$ and calculate the \emph{neighbor-average} pulse vector
\begin{align}\label{eq:neighbor-average}
    \widehat{\param_{i}} = \frac 1 {|\eta(p_i)|} \sum_{p_j \in \eta(p_i)} \param_j,
\end{align}
which is the average over the pulse variables $\param_j$ of all neighboring points $p_j$. We then calculate the Tikhonov penalty between the optimized pulse $\param_i$ and the neighbor-average pulse $\widehat{\param_{i}}$. An example of a neighbor-average pulse calculation is shown in Figure \ref{fig:example}c.

For each point (ordered by highest neighbor-average Tikhonov penalty), we recalculate $\widehat{\param_{i}}$ and then use the optimal control solver to re-optimize $\param_i$ using cost function \eqref{eq:cost}, with initial guess and $\param_0$ both set to be $\widehat{\param_{i}}$. The Tikhonov regularization term encourages the optimized pulse $\param_i$ to be as similar as possible to $\widehat{\param_i}$ while still yielding low infidelity.

This tune-up procedure can be repeated multiple times, with each round of re-optimization building upon the last and steering the reference pulses to be increasingly similar to each other.

\section{Interpolating between reference pulses}\label{sec:interpolation}

Once the reference points have been calibrated, to calculate the linearly-interpolated control pulse for some new point $\widetilde p$, we construct a weighted sum of reference pulses $\param_i$ at nearby reference points $\{p_i\}$. We first locate the point within one of the simplices in the Delaunay mesh (the same mesh we previously used in the calibration process). The interpolated pulse vector is then a linear combination of the reference pulses at the vertices of this simplex, weighted by barycentric coordinates. 

Barycentric coordinates are a group of $d+1$ numbers that uniquely identify a point within a $d$-dimensional simplex, each corresponding to one vertex of the simplex. Barycentric coordinates are uniquely determined by the requirement that the target point is equal to the coordinate-weighted sum of the vertices,
\begin{align}
    \widetilde p = \sum_{p_i \in S_{\widetilde p}} b_i p_i,
\end{align}
where $b_i$ is the barycentric coordinate of $\widetilde p$ with respect to vertex $p_i$ and $S_{\widetilde p}$ is the set of vertices defining the simplex that contains point $\widetilde p$. The closer the point is to a given vertex, the larger the corresponding coordinate will be \footnote{If the point lies on a face of the simplex, the only nonzero barycentric coordinates will be those corresponding to the vertices that define that face. This property ensures that interpolations are consistent along face boundaries of adjacent simplices. Along with the fact that barycentric coordinates vary continuously within a simplex, this means that the interpolation space is continuous throughout the simplicial mesh.}. Each coordinate is bounded by $[0,1]$ and the coordinates sum to $1$. 

Given these coordinates, our interpolated pulse vector is determined by
\begin{align} \label{eq:interpolation}
    \param_{\widetilde p} = \sum_{p_i \in S_{\widetilde p}} b_i \param_i
\end{align}
where the vector $\param_i$ is the optimized reference pulse at vertex $p_i$. This barycentric coordinate approach is the generalization of linear interpolation to arbitrary dimension. An example of this interpolation process in three dimensions is shown in Figure \ref{fig:example}d.

\section{Steps of general method}

In this section, we present our general approach to calibrate a continuous set of quantum control pulses.

\begin{enumerate}
    \item \emph{Setup.} If using a model-based optimizer, obtain a model of the device, such as a Hamiltonian. Choose a pulse description $\param$ (a finite set of variables used to construct each pulse) and a pulse optimization algorithm. Define the parameters of the gate family and determine the space to interpolate within. Sample a number of parameter points $\{p_i\}$ from this space and obtain the corresponding quantum operations. 

    \item \emph{Initial optimization.} Use the optimizer to generate initial reference pulses $\param_i$ for each reference point.

    \item \emph{Re-optimization.} For each reference point $p_i$: 
    
    \begin{enumerate}
        \item Calculate new target pulse $\param_{0,i}$ based on the set of existing reference pulses.
        
        \item Re-optimize the reference pulse $\param_{i}$ with this target pulse as the initial guess. Use Tikhonov regularization in the pulse optimization cost function to encourage the final pulse to be close to the target pulse.
    \end{enumerate} 

    Repeat step 3 as needed.

    \item \emph{Interpolation.} Choose an interpolation function $f:(\widetilde p, \{p_i, \param_i\}) \to \param_{\widetilde p}$ that calculates interpolated pulse $\param_{\widetilde p}$ at parameter-space point $\widetilde p$ given the set of optimized reference points and pulses $\{p_i, \param_i\}$.
    
\end{enumerate}

\section{Example: the Weyl chamber of two-qubit gates}\label{sec:evaluation}

We demonstrate this re-optimization and interpolation scheme for the three-parameter Cartan decomposition of two-qubit gates. Any two-qubit quantum logic gate $U \in SU(4)$ can be written as
\begin{equation} \label{eq:cartan}
    U = k_1 \exp\Bigg(-i \frac \pi 2\sum_{j=x,y,z} t_j \sigma_j^{(1)} \sigma_j^{(2)}\Bigg) k_2
\end{equation}
in terms of Pauli matrices $\sigma_x,\sigma_y,\sigma_z$ and Cartan coordinates $t_x, t_y, t_z$ \cite{zhang_geometric_2003}. The operations $k_1, k_2 \in SU(2) \otimes SU(2)$ represent single-qubit gates acting on the qubits independently, i.e. $k_1 = U_0 \otimes U_1$ for some single-qubit gates $U_0$ and $U_1$. Distinct two-qubit gates are referred to as ``locally equivalent'' or ``equivalent up to single-qubit gates'' if they have the same Cartan coordinates.

\begin{figure}[t]
    \centering
    \includegraphics[width=\linewidth]{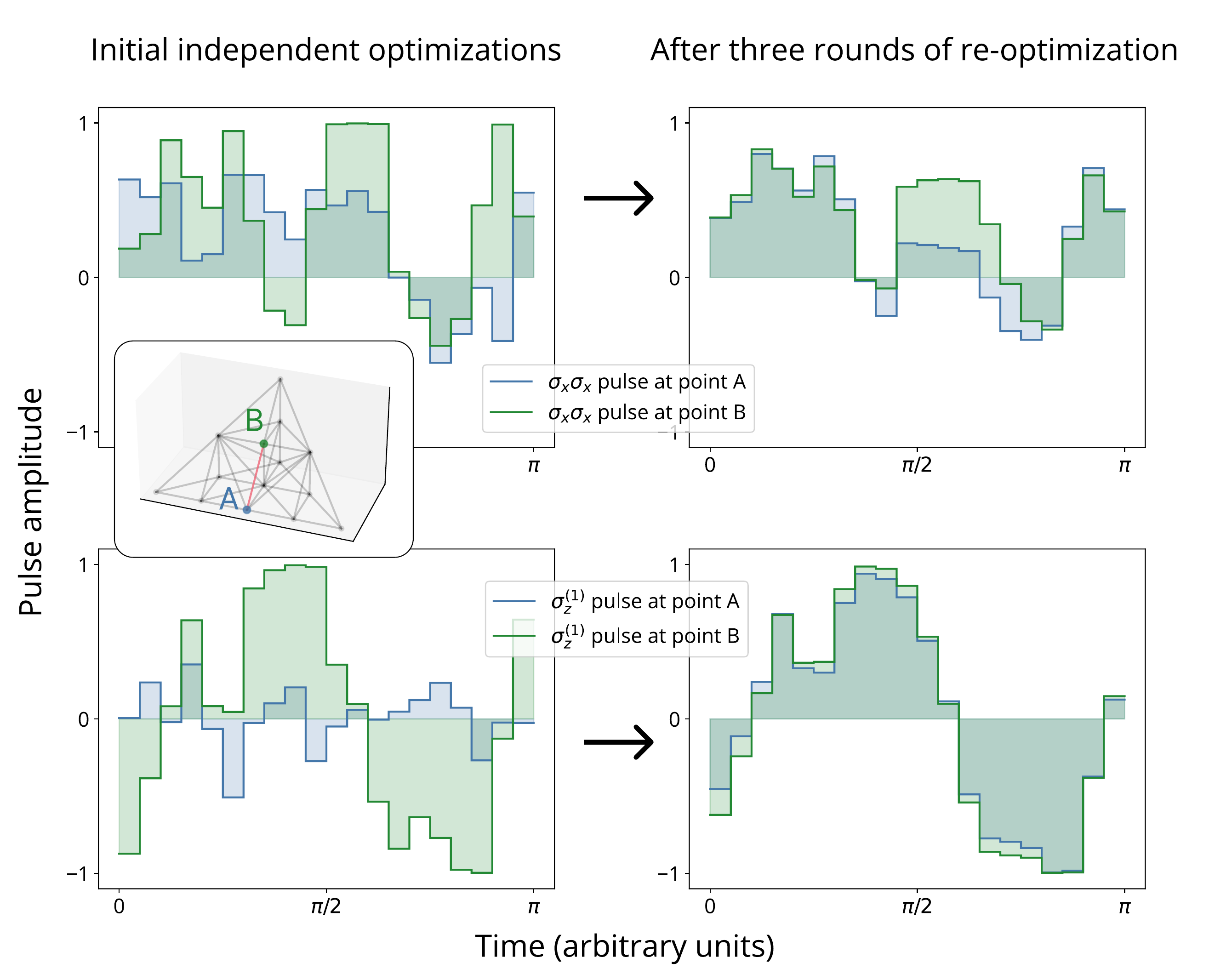}
    \caption{Comparison of pulse shapes for two adjacent reference points before and after several rounds of re-optimization. Point A corresponds to Cartan coordinates $(\frac 1 2, 0, 0)$ (the coordinates of $CNOT$ and $CZ$) and point B corresponds to coordinates $(\frac 1 2, \frac 1 4, \frac 1 4)$. Optimized controls for the two operations (from the same results as shown in Figure \ref{fig:interpolation_comparison}) are shown for the $\sigma_x \sigma_x$ control (top) and $\sigma_z^{(1)}$ control (bottom) of the Hamiltonian \eqref{eq:hamiltonian}. \emph{Inset}: locations of points A and B in the Weyl chamber. \emph{Left}: the pulse shapes are initially significantly different between points A and B. \emph{Right}: the pulses become far more similar after re-optimization, making interpolation easier, but still retain certain differences in their shapes that account for the differences in the resulting operations.}
    \label{fig:pulses}
\end{figure}

The Weyl chamber is a volume in Cartan coordinate space defined by the equations
\begin{align} \label{eq:coords}
    0 &\leq t_x \leq 1\notag\\
    0 &\leq t_y \leq \min(t_x, 1-t_x)\\
    0 &\leq t_z \leq t_y\notag.
\end{align}
The Weyl chamber is shown in Figure \ref{fig:example}a. It contains the Cartan coordinates of every two-qubit gate, making it a compelling example to demonstrate our interpolation method; the ability to perform any operation in the Weyl chamber with high fidelity using a single pulse could significantly improve the performance of existing quantum computers. For a more in-depth introduction to Cartan coordinates we refer the reader to \cite{crooks_gates_2022}. We note that pulse optimization for target operations within the Weyl chamber has been considered before, e.g. \cite{muller_optimizing_2011, watts_optimizing_2015, goerz_optimizing_2015}; however, here we demonstrate optimization for \emph{all} operations in the Weyl chamber together, which to the best of our knowledge has only been considered in \cite{sauvage_optimal_2022}. In this work, we demonstrate our interpolation scheme for operations of the form \eqref{eq:cartan} with $k_1=k_2=I$ for all parameter points that lie within the Weyl chamber \eqref{eq:coords}.

To evaluate the performance of our approach in comparison to \cite{sauvage_optimal_2022}, we use the same two-qubit Hamiltonian
\begin{align}\label{eq:hamiltonian}
    H (t) &= f^{\param}_{xx}(t) \sigma_x^{(1)} \sigma_x^{(2)} + \sum\limits_{j=1}^2 f^{\param}_{jy}(t) \sigma_y^{(j)} + f^{\param}_{jz}(t) \sigma_z^{(j)}
\end{align}
for specific parameter values $\param$. The five control functions $f(t)$ are each restricted to values in $[-1,1]$, and the pulse duration is fixed to $\pi$. We optimize reference pulses using the procedures described previously and then test the interpolation quality by evaluating the infidelities of interpolated pulses for new operations in the space. We perform several rounds of neighbor-average re-optimization and track the improvement in average infidelity.

Each control function $f(t)$ is decribed as a piecewise-constant function of 20 segments, yielding 100 total optimizable parameters. We set the Tikhonov regularization weight $\lambda = 10^{-2}$. The pulse optimizer is run for a maximum of 50 iterations for each individual optimization (or re-optimization), although it often terminates early upon reaching convergence, especially for later rounds of re-optimization where the pulse shapes do not change as drastically. We found this iteration limit to be sufficient for convergence over multiple re-optimization rounds, and it also avoids unnecessary computation during early rounds of optimization (where the pulse shapes need not be finalized). We use the default Q-CTRL convergence criteria \cite{ball_software_2021}.

\begin{figure}[t]
    \centering
    \includegraphics[width=\linewidth]{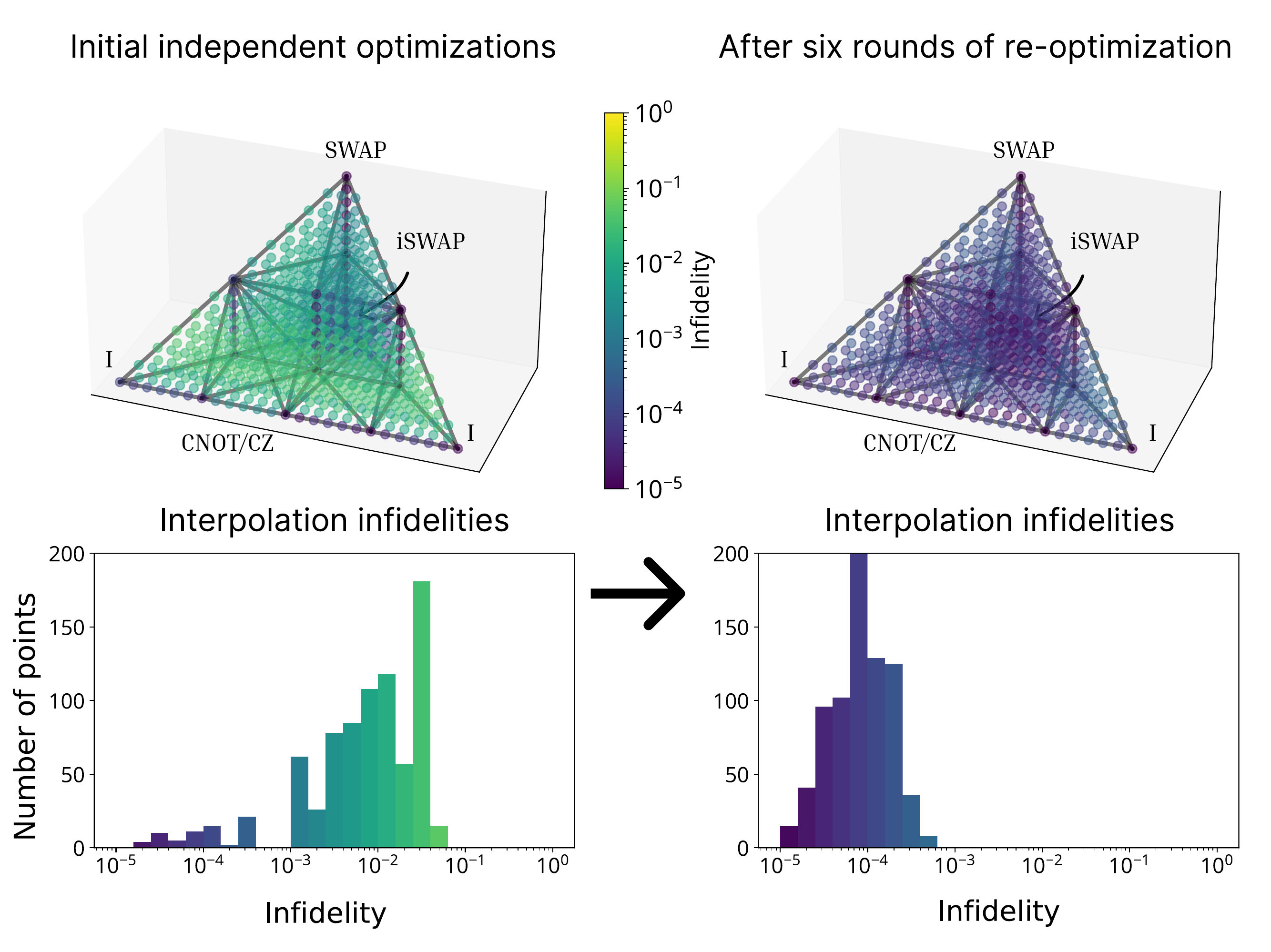}
    \caption{Interpolation quality after applying the re-optimization scheme. Axes on the top plots correspond to the three Cartan coordinates of two-qubit gates. Reference pulses are generated by optimal control software for 14 reference points in parameter space. Pulses for any point in the chamber can be obtained by linear interpolation between reference pulses. \emph{Left}: interpolated pulse infidelities at 819 test points (granularity 1/24) after initial optimization of reference points. Black lines connect reference points and define the simplicial mesh that is used for interpolation. Each colored point represents a unitary operation defined by its coordinates. The color of each point is the infidelity of the pulse that is obtained via interpolation. Mean infidelity is $1.3 \pm 1.2 \times 10^{-2}$. \emph{Right}: results after repeatedly re-optimizing each reference point to be near the average of its neighbors. Mean infidelity is improved to $1.0 \pm 0.8 \times 10^{-4}$.}
    \label{fig:interpolation_comparison}
\end{figure}

\begin{figure}[t]
    \centering
    \includegraphics[width=\linewidth]{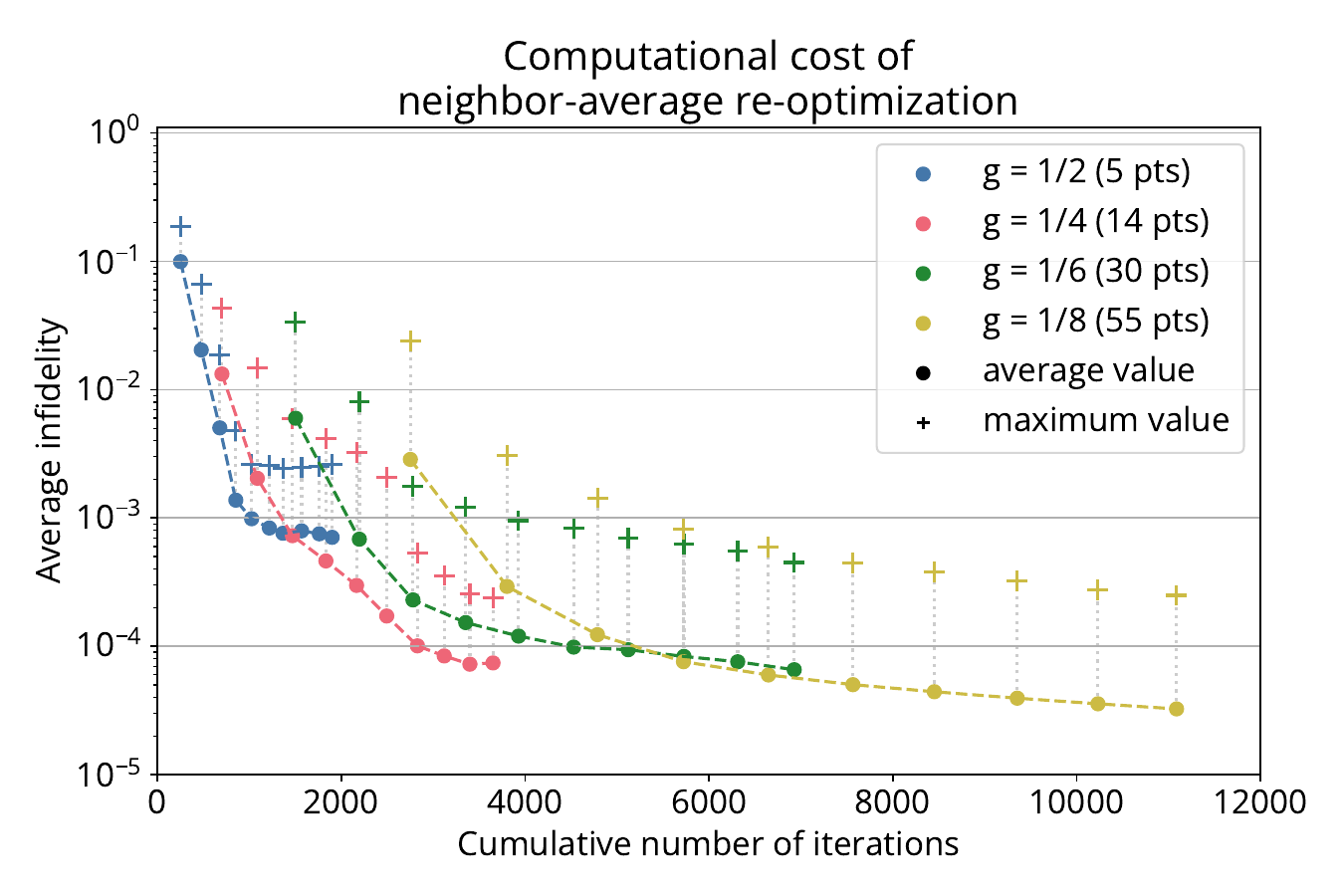}
    \caption{Different reference point granularities translate to varying amounts of classical computation time needed to optimize all reference pulses. Consecutive points with the same granularity $g$ correspond to subsequent re-optimization rounds, which can further improve average (and maximum) infidelity at the cost of more optimizer iterations. Each optimization iteration corresponds to one system evolution. The points indicated by the plus sign indicate the worst infidelity of any test point.}
    \label{fig:computation}
\end{figure}

In Figure \ref{fig:pulses} we examine two specific control pulses for two different reference points in the Weyl chamber. The two points of interest are connected by an edge in the simplicial mesh, and thus are influenced by each other in the neighbor-average re-optimization process. Before the re-optimization steps, the pulse shapes are noticably different, which leads to lower-quality interpolation in between the pulses, as can be seen in the left side of Figure \ref{fig:interpolation_comparison}. After re-optimization, the pulse shapes are much more similar, although some differences remain (the pulses cannot be identical because they perform different operations). The infidelity at the test point directly between these two reference points improved from $3.9 \times 10^{-2}$ initially to $1.5 \times 10^{-4}$ after the final round of re-optimization. In the right side of Figure \ref{fig:interpolation_comparison}, we observe a significant improvement in interpolated pulse accuracy across the entire Weyl chamber.

\begin{figure*}[t]
    \centering
    \includegraphics[width=\textwidth]{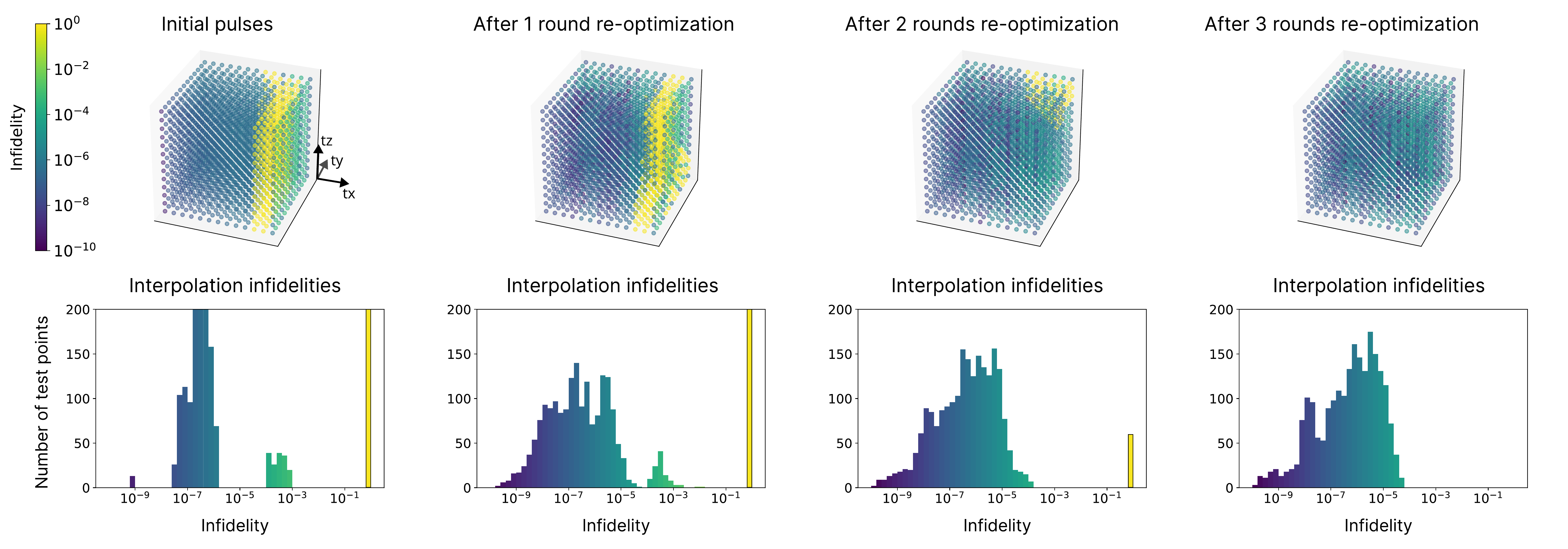}
    \caption{Infidelities at 2197 test points within the parameter space of single-qubit rotations (see Equation \eqref{eq:single-qubit}). Average and worst-case infidelities are shown in Table \ref{tab:sq-infidelities}. We observe a region of extremely poor interpolation quality (yellow), indicating that the pulses on either side of this area have very different shapes and actions. Intuitively, the pulses on the two different sides of this boundary take different routes around the Bloch sphere to reach the target unitary. Subsequent rounds of neighbor-average re-optimization reduce the size of this poor interpolation region and eventually eliminate it, resulting in an average interpolation infidelity (across all test points) of $3.5 \pm 6.6 \times 10^{-6}$ after three rounds of re-optimization.}
    \label{fig:sq-comparison}
\end{figure*}

\begin{table*}[t]
    \renewcommand{\arraystretch}{1.3}
    \caption{Interpolation infidelities for the space of single-qubit rotations after varying rounds of re-optimization.}
    \begin{tabular*}{\linewidth}{@{\extracolsep{\fill}} lllll|l}
         \hline
         Re-optimization rounds & 0 & 1 & 2 & 3 & Ref. \cite{sauvage_optimal_2022}\\
         \hline
         Average infidelity & $1.3(2.8) \times 10^{-1}$ & $1.2(2.7) \times 10^{-1}$ & $2.0(12) \times 10^{-2}$ & $\mathbf{3.5(6.6) \times 10^{-6}}$ & $4(5) \times 10^{-4}$\\
         Maximum infidelity & $7.7 \times 10^{-1}$ & $7.8 \times 10^{-1}$ & $7.7 \times 10^{-1}$ & $\mathbf{5.4 \times 10^{-5}}$ & N/A\\
         Total iterations & $4680$ & $5445$ & $6072$ & $\mathbf{6654}$ & $51200$\\
         \hline
    \end{tabular*}
    
    \label{tab:sq-infidelities}
\end{table*}

\begin{table*}[t]
    \renewcommand{\arraystretch}{1.3}
    \caption{Interpolation infidelities for the Cartan coordinate $[0,1]^3$ box after varying rounds of re-optimization.}
    \begin{tabular*}{\linewidth}{@{\extracolsep{\fill}} llllll|l}
        \hline
         Re-optimization rounds & 0 & 1 & 2 & 3 & 4 & Ref. \cite{sauvage_optimal_2022}\\
         \hline
         Average infidelity & $1.2(2.5) \times 10^{-1}$ & $8.2(21) \times 10^{-2}$ & $4.6(16) \times 10^{-2}$ & $6.6(60) \times 10^{-3}$ & $\mathbf{4.2(6.6) \times 10^{-4}}$ & $4(4) \times 10^{-4}$\\
         Maximum infidelity & $7.1 \times 10^{-1}$ & $6.8 \times 10^{-1}$ & $6.7 \times 10^{-1}$ & $6.1 \times 10^{-1}$ & $\mathbf{3.9 \times 10^{-3}}$ & N/A\\
         Total iterations & $6270$ & $11068$ & $15157$ & $18406$ & $\mathbf{21377}$ & $51200$\\
         \hline
    \end{tabular*}
    
    \label{tab:cartan-infidelities}
\end{table*}

To investigate the relationship between calibration cost and accuracy, we evaluate the performance of our method for several different reference point granularities and varying numbers of re-optimization rounds, which both influence the total computation time. We evaluate reference point granularities from $\frac 1 2$ to $\frac 1 8$ and perform up to 10 rounds of re-optimization. We use the cumulative sum of all iterations used by the optimizer (equivalent to the total number of cost function evaluations) as a measurement of the classical computation time required to calibrate the interpolation landscape. Our results are shown in Figure \ref{fig:computation}.  We find that performance generally improves with more computation, whether this is achieved by more rounds of re-optimization or a denser sampling of reference points. However, we observe diminishing returns when applying many rounds of re-optimization at the same reference point granularity; simply performing more rounds of re-optimization will not always yield better results, and moving to a higher density of reference points can give a much better performance gain for the same amount of total computation time. Overall, the data suggests that computation time is a relatively accurate predictor of interpolation quality regardless of the specific reference point distribution or number of re-optimization rounds.

\section{Comparison to previous methods}

Finally, we evaluate our approach in comparison to existing methods. We compare to \cite{sauvage_optimal_2022} for two gate families. The first is the three-parameter family of single-qubit gates
\begin{equation} \label{eq:single-qubit}
    U = \exp\bigg(-i \frac \pi 2 \Big(t_x \sigma_x + t_y \sigma_y + t_z \sigma_z\Big)\bigg)
\end{equation}
for $t_x,t_y,t_z \in [0,1]$. The second family is the three-parameter family of all two-qubit gates that we have studied in previous sections \eqref{eq:cartan}. However, instead of restricting to the Weyl chamber, \cite{sauvage_optimal_2022} considers a larger volume within this space, specifically $t_x,t_y,t_z \in [0,1]$. This space is 24 times larger than the Weyl chamber, making it significantly more computationally expensive to calibrate using our method\footnote{The Weyl chamber alone contains the Cartan coordinates of all two-qubit gates, so this larger space merely serves to increase the complexity of the problem without adding any additional benefit.}. We use the same Hamiltonian model \eqref{eq:hamiltonian} (considering only the first qubit for the single-qubit gate case). 
% For both cases, Ref. \cite{sauvage_optimal_2022} uses 51,200 cost function evaluations in total (400 training iterations with 128 sample points per iteration) and achieves average infidelities of $4 \pm 5 \times 10^{-4}$ for the space of all single-qubit gates and $4 \pm 4 \times 10^{-4}$ for the space of all two-qubit gates.

\subsection{Results}

For the single-qubit gate family, \cite{sauvage_optimal_2022} reports average pulse infidelity of $4 \pm 5 \times 10^{-4}$ after using 51,200 system evolutions to train the neural network. Our results for the same space of single-qubit gates are shown in Figure \ref{fig:sq-comparison} for various numbers of re-optimization rounds. We use a reference point granularity of 1/4 (125 points). Infidelities are evaluated on a grid in parameter space with granularity 1/12 (2197 test points). Large regions of parameter space appear to interpolate reasonably well after the initial optimization round; we attribute this to the relatively simple operation space, and to the initial Tikhonov regularization that rewards low-amplitude pulses. As shown in Table \ref{tab:sq-infidelities}, our method reaches an average pulse infidelity of $3.48 \pm 6.63 \times 10^{-6}$ in 6,654 total iterations (where each iteration corresponds to a system evolution), an improvement of over two orders of magnitude in average infidelity with 7.7x lower calibration cost compared to \cite{sauvage_optimal_2022}.

For the two-qubit case, \cite{sauvage_optimal_2022} achieves an average infidelity of $4 \pm 4 \times 10^{-4}$ using 51,200 system evolutions. We use a reference point granularity of 1/6 (343 points) and once again test interpolation quality on 2197 points. Our results for the two-qubit Cartan coordinate space are summarized in Table \ref{tab:cartan-infidelities}. We obtain a similar average infidelity to \cite{sauvage_optimal_2022} after two rounds of re-optimization with 2.4x lower calibration cost.

We observe a sharp jump in average infidelity between rounds 2 and 3 in the single-qubit case (Table \ref{tab:sq-infidelities}), whereas the two-qubit example shows a much more consistent downwards trend, improving by around 1 order of magnitude after each round of re-optimization. This is caused by the yellow region that is visible in the testing data in Figure \ref{fig:sq-comparison}; the very large infidelities at these points significantly affect the overall average. Once this yellow region disappears, the maximum test point infidelity and the average infidelity both decrease by four orders of magnitude.

We also observe a relatively small additional iteration cost of performing subsequent rounds of re-optimization. We attribute this to the fact that it takes fewer iterations for the pulse optimizer to find a good solution when it starts from a good initial guess.

These results demonstrate that our method can obtain similar or better results than previous methods with lower computational costs. We expect that calibration cost of our method could be further reduced in several simple ways. For example, the infidelity convergence threshold of the pulse optimizer could be reduced for intermediate re-optimization rounds, since the reference pulses are re-optimized several times and intermediate results are therefore not required to be perfect. Additionally, some reference points may not need to be re-optimized as many times as others if they already have very low neighbor-average Tikhonov penalty.

% \begin{table*}[t]
%     \caption{Weyl chamber interpolation infidelities after three rounds of neighbor-average re-optimization, averaged over 819 test points. Results for smaller numbers of rounds are shown in Figure 3 of the Main Text.}
%     \begin{tabular*}{\linewidth}{@{\extracolsep{\fill}} ccccc}

%          Reference point granularity & Number of points in Weyl chamber & Average infidelity & Maximum (worst) infidelity & Total iterations\\
%          \hline
%          1/2 & 5 & $5.83(2.63) \times 10^{-3}$ & $2.12 \times 10^{-2}$ & 867\\
%          1/4 & 14 & $4.88(5.53) \times 10^{-4}$ & $2.81 \times 10^{-3}$ & 2055\\
%          1/6 & 30 & $1.47(1.96) \times 10^{-4}$ & $1.23 \times 10^{-3}$ & 4005\\
%          1/8 & 55 & $6.86(9.59) \times 10^{-5}$ & $4.91 \times 10^{-4}$ & 6672
%     \end{tabular*}
    
%     \label{tab:weyl-infidelities-SI}
% \end{table*}

% \begin{table}[t]
%     \caption{Point granularities used in examples and corresponding number of points.}
%     \begin{tabular*}{\linewidth}{@{\extracolsep{\fill}} c c c}
%          \makecell{Point\\granularity} & \makecell{Number of points\\in Weyl chamber} & \makecell{Number of points\\in $[0,1]^3$ box}\\
%          \hline
%          1/2 & 5 & 27\\
%          1/4 & 14 & 125\\
%          1/6 & 30 & 343\\
%          1/8 & 55 & 729 \\
%          1/12 & 140 & 2197\\
%          1/16 & 285 & 4913\\
%          1/24 & 819 & 15625
         
%     \end{tabular*}
    
%     \label{tab:granularities-SI}
% \end{table}

\section{Alternative approaches to optimization and interpolation}

In this paper, we demonstrate our procedure using the neighbor-average Tikhonov regularization method and piecewise-linear interpolation. However, the general idea of creating reference pulses to interpolate between could have many possible implementations. Below, we list several possible modifications of our method.

\begin{itemize}
    \item \emph{Nonuniform reference point distribution}. Reference points need not be distributed on a rectangular grid; the Delaunay mesh methods will work just as well with a nonuniform distribution of points. It may be effective to use more densely-packed reference points in certain parts of parameter space to give better interpolations, or less dense points in other areas to improve calibration efficiency.

    \item \emph{Different guess methods for Tikhonov re-optimization}. In this work, we use the simple neighbor-average method to generate the initial guesses for our Tikhonov regularization re-optimization approach. However, our general approach could also be applied using guesses generated in some other way. For example, each new reference point guess could be a weighted sum of all reference pulses, with weights determined by the trace distance between the unitary operations.

    \item \emph{Selective re-optimization}. Some reference points likely do not need to be re-optimized if they already have a high degree of similarity to neighboring reference pulses; computation time could be easily improved by only re-optimizing the points that differ from nearby pulses by more than some threshold amount. This would also provide a precise stopping criteria for the calibration process.

    \item \emph{Higher-order interpolation fits}. Regardless of optimization method, different methods can be used for the interpolation step of the procedure. In this paper, we show results using a simple piecewise-linear interpolation. Alternatively, interpolation could be done using a higher-order model such as a spline or other function.
    
    %optimizations\item \emph{Incremental reference re-seeding}. Optimize each reference point for some small number of iterations $k$. Once all reference points have been optimized to $k$ iterations, begin a new round of optimizations for another $k$ iterations, starting where the previous round left off, but with updated Tikhonov target pulses. This is similar to the previous idea of reference re-seeding but without waiting for the optimizations to fully converge before updating Tikhonov targets.
    
    % \item \emph{Minimal-curvature optimization and interpolation}. Instead of running individual optimizations at each reference point, all reference pulses could be tuned at once in one large optimization. We imagine a cost function consisting of the standard infidelity term, averaged over all reference points, as well as a new term that is proportional to the total curvature (or some other metric) of an interpolation function fitted to the reference point pulses. The interpolating function would then be built into the optimization directly, potentially yielding more accurate results.
\end{itemize}

We do not perform these other interpolation methods for this paper due to the effectiveness of the simpler linear model and lack of a clearly-motivated choice of a more intricate method. However, our general framework is flexible and will work with any optimization or interpolation method, and we anticipate that more complex methods could potentially yield better results, depending on the system and gate family of interest.

\section{Discussion}

Our method improves on the work of \cite{sauvage_optimal_2022} by achieving similar pulse accuracy with less total computation required. The procedure has the additional benefits of explainability and modularity. The interpolation space contains known reference points with high-fidelity pulses, with the benefit that specific points can be re-optimized without needing to recalibrate the entire space. Modularity is important in the case where simple optimal control pulses do not transfer well to experimental settings (which is generally expected due to significant device variations over time \cite{bylander_noise_2011, chan_assessment_2018, klimov_fluctuations_2018}); more advanced optimization routines such as data-driven or robust methods \cite{leung_robust_2018, valahu_quantum_2022, wu_data-driven_2018, brown_arbitrarily_2004, goldschmidt_model_2022} can directly replace the optimal control unit without needing to redesign the rest of the procedure.

The Weyl chamber example demonstrates the potential of this calibration routine, allowing the pulses for any two-qubit gates (up to single-qubit operations) to be instantly obtained with infidelities of $1.0 \pm 0.8 \times 10^{-4}$ after only 2828 iterations of the pulse optimizer. On a real device, a similar calibration routine could allow quantum hardware to natively support any two-qubit operation directly at the pulse level, which could yield significant improvements in both execution time and accuracy by avoiding expensive gate decompositions.

We have described a method to generate a continuous space of control pulses which are constrained to be as similar as possible to each other. In addition to integrating robust pulse optimization methods, it may be beneficial to recalibrate this control pulse space over time to account for device variation over time, which is a significant problem in current quantum hardware \cite{bylander_noise_2011, chan_assessment_2018, klimov_fluctuations_2018}. We suspect that this recalibration could be significantly less expensive than simply recalibrating every reference pulse due to the high degree of similarity among nearby pulses (e.g. Figure \ref{fig:pulses}).

Our neighbor-average re-optimization procedure makes no assumptions about the structure of the parameter space of interest, and thus may not be optimal; we expect that better-motivated approaches could potentially yield higher-accuracy or more efficient interpolations by taking into account the structure of the gate family of interest. The specific method we use in this paper is easily-implemented and conceptually simple, so we provide it as a baseline for future improvements.

Python code used to generate the results in this work is publicly available in a Github repository \cite{chadwick_pulse-interpolation_2023}.

We thank Lennart Maximilian Seifert and Andy Goldschmidt for helpful discussions on quantum control and the interpolation scheme, and Andr\'e Carvalho and the Q-CTRL team for enabling our use of their software tools in this paper. We also thank the anonymous QCE reviewers for their helpful comments. We used the ``Bright'' colorblind-friendly plot color scheme from \cite{tol_qualitative_2021}. This work is funded in part by EPiQC, an NSF Expedition in Computing, under award CCF-1730449; in part by STAQ under award NSF Phy-1818914; in part by the US Department of Energy Office of Advanced Scientific Computing Research, Accelerated Research for Quantum Computing Program; and in part by the NSF Quantum Leap Challenge Institute for Hybrid Quantum Architectures and Networks (NSF Award 2016136) and in part based upon work supported by the U.S. Department of Energy, Office of Science, National Quantum Information Science Research Centers. FTC is Chief Scientist for Quantum Software at Infleqtion and an advisor to Quantum Circuits, Inc.

\bibliography{references.bib}

\end{document}

\title{Supplemental Material:\\Efficient control pulses for continuous quantum gate families\\through coordinated re-optimization}

\author{Jason D. Chadwick}
 \email{jchadwick@uchicago.edu}
\author{Frederic T. Chong}
\email{chong@cs.uchicago.edu}
\affiliation{
 Department of Computer Science, University of Chicago, Chicago, IL 60637, USA
}

\date{\today}

%\keywords{Suggested keywords}%Use showkeys class option if keyword

\maketitle

All example results used in the Main Text as well in this Supplemental Material are obtained using Python code that is available on Github \cite{chadwick_pulse-interpolation_2023}. A valid Q-CTRL license is needed to fully reproduce the results from scratch.

\begin{table*}[t]
    \caption{Weyl chamber interpolation infidelities after three rounds of neighbor-average re-optimization, averaged over 819 test points. Results for smaller numbers of rounds are shown in Figure 3 of the Main Text.}
    \begin{tabular}{c c c c}
         Reference point granularity & Average infidelity & Maximum (worst) infidelity & Total iterations\\
         \hline
         1/2  & $5.83(2.63) \times 10^{-3}$ & $2.12 \times 10^{-2}$ & 867\\
         1/4 & $4.88(5.53) \times 10^{-4}$ & $2.81 \times 10^{-3}$ & 2055\\
         1/6 & $1.47(1.96) \times 10^{-4}$ & $1.23 \times 10^{-3}$ & 4005\\
         1/8 & $6.86(9.59) \times 10^{-5}$ & $4.91 \times 10^{-4}$ & 6672
         
    \end{tabular}
    
    \label{tab:weyl-infidelities-SI}
\end{table*}

\begin{table}[t]
    \caption{Point granularities used in examples and corresponding number of points.}
    \begin{ruledtabular}
    \begin{tabular}{c c c}
         \makecell{Point\\granularity} & \makecell{Number of points\\in Weyl chamber} & \makecell{Number of points\\in $[0,1]^3$ box}\\
         \hline
         1/2 & 5 & 27\\
         1/4 & 14 & 125\\
         1/6 & 30 & 343\\
         1/8 & 55 & 729 \\
         1/12 & 140 & 2197\\
         1/16 & 285 & 4913\\
         1/24 & 819 & 15625
         
    \end{tabular}
    \end{ruledtabular}
    
    \label{tab:granularities-SI}
\end{table}

\begin{table*}[t]
    \caption{Interpolation infidelities for the Cartan coordinate $[0,1]^3$ box after varying rounds of re-optimization.}
    \begin{ruledtabular}
    \begin{tabular}{c c c c}
         \makecell{Number of\\re-optimization rounds} & Average infidelity & Maximum (worst) infidelity & Cumulative total iterations\\
         \hline
         0  & $3.27(16.2) \times 10^{-2}$ & $1.00 \times 10^{0}$ & 17213\\
         1 & $6.62(16.1) \times 10^{-4}$ & $1.95 \times 10^{-2}$ & 26631\\
         2 & $3.55(6.72) \times 10^{-4}$ & $6.54 \times 10^{-3}$ & 34407\\
         3 & $2.54(4.33) \times 10^{-4}$ & $3.91 \times 10^{-3}$ & 41289
         
    \end{tabular}
    \end{ruledtabular}
    
    \label{tab:cartan-infidelities-SI}
\end{table*}

\section{Other approaches to reference optimization and interpolation}\label{sec:other-approaches}

In this paper, we demonstrate our procedure using the neighbor-average Tikhonov regularization method and simple piecewise-linear interpolation. However, the general idea of creating reference pulses to interpolate between could have many possible implementations. Below, we list several alternative ways of realizing this methodology.

\begin{itemize}
    \item \emph{Nonuniform reference point distribution}. Reference points need not be distributed on a rectangular grid; the Delaunay triangulation method will work just as well with a nonuniform distribution of points. It may be effective to use more densely-packed reference points in certain parts of parameter space to give better interpolations, or less dense points in other areas to improve calibration efficiency.

    \item \emph{Different guess methods for Tikhonov re-optimization}. We used the simple neighbor-average method to generate the initial guesses for our Tikhonov regularization re-optimization approach. However, our general approach could also be applied using guesses generated in some other way. For example, each new reference point guess could be a weighted sum of all reference pulses, with weights determined by some notion of distance between the operations.

    \item \emph{Selective re-optimization}. Some reference points likely do not need to be re-optimized if they already have a high degree of similarity to neighboring reference pulses; computation time could be easily improved by only re-optimizing the points that differ from nearby pulses by more than some amount.

    \item \emph{Higher-order interpolation fits}. Regardless of optimization method, different methods can be used for the interpolation step of the procedure. In this paper, we show results using a simple piecewise-linear interpolation. Alternatively, interpolation could be done using a higher-order model such as a spline or other function.
    
    %optimizations\item \emph{Incremental reference re-seeding}. Optimize each reference point for some small number of iterations $k$. Once all reference points have been optimized to $k$ iterations, begin a new round of optimizations for another $k$ iterations, starting where the previous round left off, but with updated Tikhonov target pulses. This is similar to the previous idea of reference re-seeding but without waiting for the optimizations to fully converge before updating Tikhonov targets.
    
    \item \emph{Minimal-curvature optimization and interpolation}. Instead of running individual optimizations at each reference point, all reference pulses could be tuned at once in one large optimization. We imagine a cost function consisting of the standard infidelity term, averaged over all reference points, as well as a new term that is proportional to the total curvature (or some other metric) of an interpolation function fitted to the reference point pulses. The interpolating function would then be built into the optimization directly, potentially yielding more accurate results.
\end{itemize}

We do not perform these other interpolation methods for this paper due to the high effectiveness of the simpler linear model and lack of a clearly-motivated choice of a more intricate method. However, our general framework is flexible and will work with any optimization or interpolation method, and we anticipate that more complex methods could potentially yield better results (depending on the system and gate family of interest).

\section{Extension: variable pulse duration}\label{sec:duration-scaling}

In this work, we assumed that all pulses have the same fixed duration, regardless of the target quantum operation. In reality, it is much easier to perform a gate that is close to the identity than a more complex operation such as SWAP. We expect that these easier gates could be accomplished with a shorter duration within the same pulse amplitude bounds, compared to the more complex gates in the family.

Depending on the chosen pulse description, it can be possible to include the duration of the pulse directly as an optimizable quantity. The duration is then an additional element in the pulse vector and can be included in the interpolation.

Alternatively, when constrained to a specific duration as in this work, we expect that some easier operations need not make use of the full allowed amplitude of the pulse. In this case, for some Hamiltonians, it is possible to upscale the amplitude and downscale the duration equivalently to obtain a shorter pulse with the same effect. A new cost function term could be added to the optimizations to penalize the maximum value of the control field to improve the amount of downscaling possible.

\begin{figure*}[t]
    \centering
    \includegraphics[width=\textwidth]{comparison.eps}
    \caption{Infidelities at 2197 test points within the $[0,1]^3$ Cartan coordinate box for 0 to 3 rounds of neighbor-average re-optimization. Average and worst-case infidelities are shown in Table \ref{tab:cartan-infidelities-SI}.}
    \label{fig:comparison-SI}
\end{figure*}

\section{General methodology: coordinated re-optimization}

The methodology for applying our general approach (coordinated re-optimization of individual reference pulses) to an arbitrary family of quantum operations is as follows:

\begin{enumerate}
    \item \emph{Setup.} If using a model-based optimizer, obtain a model of the device, such as a Hamiltonian. Choose a pulse description $\param$ (a finite set of variables used to construct each pulse) and a pulse optimization algorithm. Define the parameters of the gate family and determine the space to interpolate within. Sample a number of parameter points $\{p_i\}$ from this space and obtain the corresponding quantum operations. 

    \item \emph{Initial optimization.} Use the optimizer to generate initial reference pulses $\param_i$ for each reference point.

    \item \emph{Re-optimization.} For each reference point $p_i$: 
    
    \begin{enumerate}
        \item Calculate some new target pulse $\param_{0,i}$ based on the set of existing reference pulses.
        
        \item Re-optimize the reference pulse with this target pulse as the initial guess. Use Tikhonov regularization in the cost function (or some other method) to encourage the final pulse to be close to the target pulse.
    \end{enumerate} 

    Repeat this step as needed.

    \item \emph{Interpolation.} Choose an interpolation function $f:(\widetilde p, \{p_i, \param_i\}) \to \param_{\widetilde p}$ that calculates interpolated pulse $\param_{\widetilde p}$ at parameter-space point $\widetilde p$ given the set of optimized reference points and pulses $\{p_i, \param_i\}$.
    
\end{enumerate}

\section{Details on Weyl chamber examples from Main Text}

For all examples, we fix the pulse duration to $\pi$ and parameterize each of the 5 control functions $f(t)$ as a piecewise-constant function of 6 segments, yielding 30 total optimizable parameters. We set the Tikhonov regularization weight $\lambda = 10^{-2}$. The pulse optimizer is run for a maximum of 50 iterations for each individual optimization (or re-optimization), although it often terminates early upon reaching convergence, especially for later rounds of re-optimization where the pulse shapes do not change as drastically. We found this to be sufficient for convergence over multiple re-optimization rounds, and it also avoids unnecessary computation during early rounds of optimization (where the pulse shapes need not be finalized). We use the default Q-CTRL convergence criteria \cite{ball_software_2021}. We use various point granularities to uniformly cover the parameter space of interest; the number of points generated by grids of various granularities (for both the Weyl chamber and the larger parameter-space box used to compare to \cite{sauvage_optimal_2022}) are shown in Table \ref{tab:granularities-SI}.

For the example used in Figures 1 and 2 in the Main Text, we use a reference point granularity of 1/4 and a test point granularity of 1/16. We initialize the reference points with naive pulse optimizations, run three consecutive rounds of neighbor-average re-optimization, and use the initial and final reference pulses to extract the data shown in the two figures. 

To demonstrate the relationship between computational cost and average infidelity (Figure 3 in the Main Text), we use reference point granularities ranging from 1/2 to 1/8 and always use test point granularity of 1/24. We run three rounds of re-optimization for each reference point granularity. After initialization and after each round of re-optimization, we record the average and maximum infidelities achieved over all test points. These values are displayed in Figure 3 of the Main Text. Table \ref{tab:weyl-infidelities-SI} shows specific values for the average and worst-case infidelities achieved after the final round of re-optimization. Increasing reference point granularity appears to significantly improve interpolation quality, but at the cost of more pulse optimizer iterations.

\section{Comparison to existing methods}

In this section, we describe our methodology for comparing the approach presented in this work with that of \cite{sauvage_optimal_2022}, which to the best of our knowledge is the only other work that addresses the problem of pulse generation for multi-parameter continuous sets of gates. It is critical to minimize the computational overhead of this problem to enable a future experimental implementation.

\subsection{Setup}

We specifically compare 3-parameter interpolation for the 2-qubit gate family described by
\begin{equation} \label{eq:cartan-SI}
    U = \exp\Bigg(-i \frac \pi 2\sum_{j=x,y,z} t_j \sigma_j^{(1)} \sigma_j^{(2)}\Bigg)
\end{equation}
where $t_x,t_y,t_z$ are the Cartan coordinates of $U$. This is the same unitary as used in the Main Text for the Weyl chamber example, but for a fair comparison with the results of Ref. \cite{sauvage_optimal_2022}, we extend the parameter bounds to $t_x,t_y,t_z \in [0,1]$. This occupies a volume 24 times larger than the Weyl chamber in parameter space. However, we note that the Weyl chamber alone can already represent any two-qubit operation (up to single-qubit operations), so this extra volume in parameter space only serves to increase computational complexity without adding any practical benefit. 

We discretize each of the five control pulses into 6 piecewise-constant segments of equal duration. As in the main paper, we use the Hamiltonian 

\begin{align}\label{eq:hamiltonian-SI}
    H (t) &= f^{\param}_{xx}(t) \sigma_x^{(1)} \sigma_x^{(2)} + \sum\limits_{j=2}^2 f^{\param}_{jy}(t) \sigma_y^{(i)} + f^{\param}_{jz}(t) \sigma_z^{(i)}
\end{align}
and set the pulse duration to $\pi$.

We pick reference points on a grid in parameter space with spacing 1/6. We find that the optimization and interpolation methods discussed in the Main Text provide good results with little need for fine-tuning.

\subsection{Results}

Interpolation results are shown in Figure \ref{fig:comparison-SI} for various numbers of re-optimization rounds. Infidelities are evaluated on a grid in parameter space with granularity 1/12. Large regions of parameter space appear to already interpolate reasonably well; we attribute this to the constraints we have imposed on the pulses, namely the small number of optimizable parameters (each pulse consisting of only 6 piecewise-constant segments) and the initial Tikhonov regularization that favors low-amplitude pulses.

Using a neural network, Ref. \cite{sauvage_optimal_2022} reports average pulse infidelity of $4 \pm 4 \times 10^{-4}$ while using 51,200 system evolutions to train the network. As shown in Table \ref{tab:cartan-infidelities-SI}, our method reaches an average pulse infidelity of $3.55 \pm 6.72 \times 10^{-4}$ in 34,407 total iterations, which can further be refined to $2.54 \pm 4.33 \times 10^{-4}$ in 41,289 total iterations. Each iteration corresponds to a system evolution.

This demonstrates that our method can obtain similar or better results than previous methods using less computation, with the additional benefits (as discussed in the Main Text) of improved explainability and modularity.

We expect that computation time could be further reduced in several simple ways, if desired. For example, the cost value convergence threshold of the pulse optimizer could be reduced, since the reference pulses are re-optimized several times and intermediate results are thus not required to be perfect. Additionally, as mentioned in the first section of this Supplemental Material, some reference points may not need to be re-optimized as many times as others.

\bibliography{references.bib}